\documentclass[superscriptaddress,twocolumn,showpacs,amsmath,amssymb]{revtex4}

\usepackage{graphicx}

\def\be{\begin{equation}}
\def\ee{\end{equation}}
\def\bea{\begin{eqnarray}}
\def\eea{\end{eqnarray}}

\renewcommand{\vec}[1]{\mbox{\boldmath $#1$}}

\begin{document}

\title{Pairing correlations and odd-even staggering 
in reaction cross sections \\ 
of weakly bound nuclei
}

\author{K. Hagino}
\affiliation{ 
Department of Physics, Tohoku University, Sendai, 980-8578,  Japan} 

\author{H. Sagawa}
\affiliation{
Center for Mathematics and Physics,  University of Aizu, 
Aizu-Wakamatsu, Fukushima 965-8560,  Japan}


\begin{abstract}
We investigate the odd-even staggering (OES) in 
reaction cross sections of weakly bound nuclei with a Glauber theory,  
taking into account the pairing correlation with the 
Hartree-Fock-Bogoliubov (HFB) method. 
We first discuss the pairing gap in extremely weakly bound nuclei 
and show that the pairing gap persists in the limit of zero separation 
energy limit even 
for single-particle orbits with the 
orbital angular momenta $l=0$ and  $l=1$. 
We then introduce the OES parameter defined as the 
second derivative of 
reaction cross sections with respect to the mass number, 
and clarify the 
relation between the magnitude of OES and the neutron separation 
energy. 
We find that the OES parameter increases considerably 
in the zero separation energy limit
for $l=0$ and $l=1$ single-particle states, while no increase is found for 
higher angular momentum orbits with {\it e.g.,} $l=3$.
We point out that the increase of OES parameter is also seen in 
the experimental reaction cross 
sections for Ne isotopes, which is well accounted for by our calculations. 
\end{abstract}

\pacs{21.10.Gv,25.60.Dz,21.60.Jz,24.10.-i}

\maketitle

\section{Introduction}

Reaction  cross sections $\sigma_R$ 
of unstable nuclei provide a powerful tool to study the structure of 
unstable nuclei such as density distribution and 
deformation \cite{Tani85,Mitt87,Ozawa00}. 
For instance, a largely extended 
structure, referred to as ``halo'', of unstable nuclei such as $^{11}$Li  \cite{Tani85}, 
$^{11}$Be \cite{Fuk91}, and $^{17,19}$C~\cite{Ozawa00} has been 
found with such measurements. 
The halo structure is 
one of the characteristic 
features of weakly bound nuclei, and has attracted lots of attention 
(see Ref. \cite{Naka09} 
for a recent discovery of halo structure in $^{31}$Ne nucleus). 

Experimentally large odd-even staggering (OES) phenomena have been 
revealed in reaction cross sections of unstable nuclei close to the 
neutron drip line, {\it e.g.,}  
in the isotopes $^{14,15,16}$C \cite{FYZ04}, $^{18,19,20}$C \cite{Ozawa00}, 
$^{28,29,30}$Ne\cite{Takechi10}, 
$^{30,31,32}$Ne \cite{Takechi10}, and $^{36,37,38}$Mg \cite{Takechi11}.
In Ref. \cite{HS11}, we have argued that 
the pairing correlations play an essential role 
in these OES.  
That is, the OES in reaction cross sections 
is intimately related to 
the so called pairing anti-halo effect discussed in Ref. \cite{Benn00}. 
On the other hand, 
there have been contradictory arguments whether the pairing gap disappears \cite{Hamamoto}
or persists\cite{ZMM11,Yamagami05}  when a nucleus reaches at 
the neutron drip line, {\it i.e.}, the single-particle energy of the 
last occupied orbit approaches the zero energy.  
If the pairing gap disappeared, the OES effect 
might be either quenched or disappeared completely, 
unless the deformation parameter is significantly different among the 
neighboring nuclei 
\cite{Minomo11}. 

In this paper, 
we first discuss the pairing correlations close to the zero energy
by the Hartree-Fock Bogoliubov (HFB) method. 
We carry out HFB calculations for the neutron 3s$_{1/2}$ orbit in $^{76}$Cr 
changing the 
separation energy in a mean field potential, 
and examine different definitions 
for an effective pairing gap parameter. 
This problem is also related with the superfluidity 
of neutron gases in the outer crust of neutron 
stars \cite{Schuck11}. 
The second motivation of this work, in addition to 
giving the details of the analysis in Ref. \cite{HS11}, 
is to propose a formula to measure the odd-even staggering 
in the reaction cross sections. 
Notice that the OES of the isotope shift 
of stable nuclei has been discussed mostly 
to clarify the deformation changes in odd-mass and even-mass 
nuclei \cite{BM75}.  
The present issue of OES in the reaction
cross sections has a similar aspect 
to the previous study in Ref. \cite{BM75} in one sense, but different in 
another aspect since it aims at studying the existence of 
the pairing correlation in nuclei close to the neutron drip line.  

The paper is organized as follows. In Sec. II, we discuss the pairing correlation 
in neutron-rich nuclei using the HFB method. 
In Sec. III, we apply a Glauber theory in order to 
calculate reaction cross sections. 
We introduce the OES parameter for reaction cross sections, and 
discuss it in relation with the pairing correlations in weakly bound 
nuclei. We then summarize the paper in Sec. IV. 

\section{Pairing gap at neutron drip line}

In the coordinate space representation,  
the HFB equations read \cite{DFT84,DNW96,B00}
\bea
 \left( \begin{array}{cc}
  \hat{h}-\lambda   &   \Delta(\vec{r})  \\
   \Delta(\vec{r})  &  -\hat{h}+\lambda  \end{array} 
   \right)
    \left( \begin{array}{c}
  u_i(\vec{r}) \\
 v_i(\vec{r})    \end{array} 
   \right)
 = E_i 
   \left( \begin{array}{c}
  u_i(\vec{r}) \\
  v_i(\vec{r})    \end{array} 
   \right),
\eea
where 
\begin{equation}
\hat{h}=-\frac{\hbar^2}{2m}\nabla^2 + V(\vec{r}), 
\end{equation}
is the mean-field Hamiltonian, $m$ being the nucleon mass. 
$V(\vec{r})$ and $\Delta(\vec{r})$ are the mean-field and the pairing potentials, 
respectively, and $E_i$ is a quasi-particle energy. 
Here, we have assumed that the
nucleon-nucleon interaction is a zero range force so that
these potentials are local.
The upper component of the pair wave function $u_i(\vec{r})$ is a non-localized 
wave function if the quasi-particle energy $E_i$ is larger than
the Fermi energy $|\lambda|$, while the lower component $v_i(\vec{r})$ is always 
localized. 
The pair potential $\Delta(\vec{r})$ in general 
has a larger surface diffuseness 
than the mean field potential $V(\vec{r})$, and goes beyond it 
due to the non-localized property of the upper component of the wave function $u_i(\vec{r})$, 
that is, due to the coupling to the continuum spectra \cite{DNW96}.  

In the mean field approximation without the pairing correlations ({\it i.e.}, 
$\Delta(\vec{r})=0$), 
the halo structure
originates from an occupation of a weakly-bound $l=0$ or $l=1$ orbit by the valence 
nucleons near the threshold~\cite{Riisager,Sagawa93}.  
The asymptotic behavior of a single particle  wave function for $s$ wave 
reads 
\begin{equation}
 \psi_i(r) \sim \exp(-\alpha_i r),
\end{equation}
where $\alpha_i$ is defined as 
$\alpha_i=\sqrt{2m|\varepsilon_i|/ \hbar ^2}$ with the Hartree-Fock (HF) energy
$\varepsilon_i$. 
The mean square radius of this wave function is then evaluated as  
\begin{equation}
 \langle r^2\rangle_{\rm HF}=\frac{\int r^2 |\psi_i(r)|^2 d\vec{r}}
{\int  |\psi_i(r)|^2 d\vec{r}} \propto 
   \frac{1}{\alpha_i^2}= \frac{\hbar^2}{2m|\varepsilon_i|},
\end{equation}
which will diverge in the limit of vanishing separation energy, 
$|\varepsilon_i| \rightarrow 0$.  
It has been shown that 
this divergence occurs not only for $s$ wave but also 
for $p$ wave, although the dependence on $|\varepsilon_i|$ is 
now $\langle r^2\rangle_{\rm HF}\propto 1/\sqrt{|\varepsilon_i|}$ 
for $l=1$ \cite{Riisager}.  

In contrast, in the presence of the pairing correlations ({\it i.e., } $\Delta(\vec{r})\neq 0$), 
the lower component of the HFB wave function, which is 
relevant to the density distribution, behaves as \cite{DNW96} 
\begin{equation}
  v_i(r) \propto \exp(-\beta_i r), 
\end{equation}
where $\beta_i$ is given
by,
\begin{equation}
\beta_i =\sqrt{\frac{2m}{\hbar ^2}(E_i-\lambda)},   
\end{equation}
using the quasi-particle energy $E_i$. 
With the canonical basis $\phi_i^{\rm (can)}(\vec{r})$ in the HFB theory, 
the quasi-particle energy 
may be approximately given by \cite{DNW96}
\begin{equation}
E_i=\sqrt{(\varepsilon_i^{\rm (can)} -\lambda)^2 +(\Delta_i^{\rm (can)})^2},
\label{EHFB} 
\end{equation}
where 
$\varepsilon_i^{\rm (can)}\equiv \langle \phi_i^{(\rm can)}|\hat{h}|\phi_i^{(\rm can)}\rangle$ and 
$\Delta_i^{\rm (can)}\equiv \langle \phi_i^{(\rm can)}|\Delta(\vec{r})|\phi_i^{(\rm can)}\rangle$. 
In the zero binding limit, $\varepsilon_i^{(\rm can)} \sim 0$ and $\lambda  \sim 0$, 
the asymptotic 
behavior of the wave function 
$v_i(r)$ is therefore determined by the gap parameter as, 
\begin{equation}
 v_i(r)\propto  \exp\left[\left(-\sqrt{\frac{2m}{\hbar^2}\Delta_i^{(\rm can)}}\right) r\right].
\label{HFB}
\end{equation}
The radius of the HFB wave function will then 
be given in the limit of small separation energy $|\varepsilon_i^{(\rm can)}|
 \rightarrow 0$ as 
\begin{equation}
 \langle r^2\rangle_{\rm HFB}
=\frac{\int r^2 |v_i(r)|^2 d{\vec{r}}}{\int  |v_i(r)|^2 d\vec{r}} \propto 
   \frac{1}{\beta_i^2} \rightarrow \frac{\hbar^2}{2m\Delta_i^{(\rm can)}}.  
\end{equation}
If  the gap parameter $\Delta_i^{(\rm can)}$ stays finite in the zero energy 
limit of $\varepsilon_i^{(\rm can)}$, 
the extremely large extension of a halo wave function 
in the HF field will be reduced substantially by the pairing correlations  
and the root-mean-square (rms) radius will not diverge. 
This is referred to as the 
anti-halo effect due to the pairing correlations \cite{Benn00}.
It was shown in Ref. ~\cite{HS11} that this is the main reason for the observed OES 
in the reactions cross sections of several drip line nuclei.

In order to study the behavior of the pairing gap in 
weakly bound nuclei, we carry out 
HFB calculations for the neutrons in $^{76}$Cr nucleus. 
To this end, we use a spherical Woods-Saxon (WS) potential, 
\begin{equation}
V(r)=V_0f(r)-\frac{V_{\rm ls}}{r}\frac{df(r)}{dr}\,\vec{l}\cdot\vec{s},
\label{Vmf}
\end{equation}
with 
\begin{equation}
f(r)=\frac{1}{1+\exp[(r-R_0)/a]}, 
\label{fermifunction}
\end{equation}
for the mean-field potential $V(r)$. 
Following Ref. \cite{YS08}, we take $V_0=-51+30(N-Z)/A$ MeV, $R_0=1.27A^{1/3}$ fm, 
$V_{\rm ls}=-0.71V_0$ MeV$\cdot$fm$^2$, and $a$=0.67 fm. 
For the HFB calculations, we use a density-dependent contact pairing 
interaction of surface type, with which the pairing potential is given by 
\begin{equation}
\Delta(r)=\frac{V_{\rm pair}}{2}\left(1-\frac{\rho(r)}{\rho_0}\right)\,\tilde{\rho}_n(r). 
\label{pairpot}
\end{equation}
Here, $\rho(r)$ and $\tilde{\rho}_n(r)$ are the total particle density 
and the neutron pairing density, 
respectively, given by 
\begin{eqnarray}
\rho(r)&=&\sum_{i=n,p}|v_i(\vec{r})|^2, \\
\tilde{\rho}_n(r)&=&-\sum_{i=n}u_i^*(\vec{r})v_i(\vec{r}).
\end{eqnarray}
We again follow Ref. \cite{YS08} and take $\rho_0$=0.16 fm$^{-3}$ and 
$V_{\rm pair}=-420$ MeV$\cdot$fm$^3$ with the energy cut off of 50 MeV above the 
Fermi energy. 
In order to construct the proton density, we use the same mean-field potential as 
in Eq. (\ref{Vmf}), but with 
$V_0=-51-30(N-Z)/A$ MeV. We also add the Coulomb potential for a uniform charge 
with a radius of $R_0$. We discretize the continuum spectra with the box boundary condition. We take the box size 
of $R_{\rm box}$=60 fm, and include the angular momentum up to $l=12$.  
Notice that 
we determine the pairing potential self-consistently 
in this model 
according to Eq. (\ref{pairpot}), 
although the Woods-Saxon potential is fixed for the mean-field part. 
The Fermi energy is also determined 
self-consistently 
according to the condition for 
the average particle number conservation, 
\be
\langle A \rangle=\int \rho(r) d\vec{r} =\int \sum_i|v_i(\vec{r})|^2 d\vec{r}. 
\ee
These self-consistencies are 
particularly important to increase the pairing gap for an extremely 
loosely bound orbit \cite{Yamagami05}.

\begin{figure} 
\includegraphics[scale=0.5,clip]{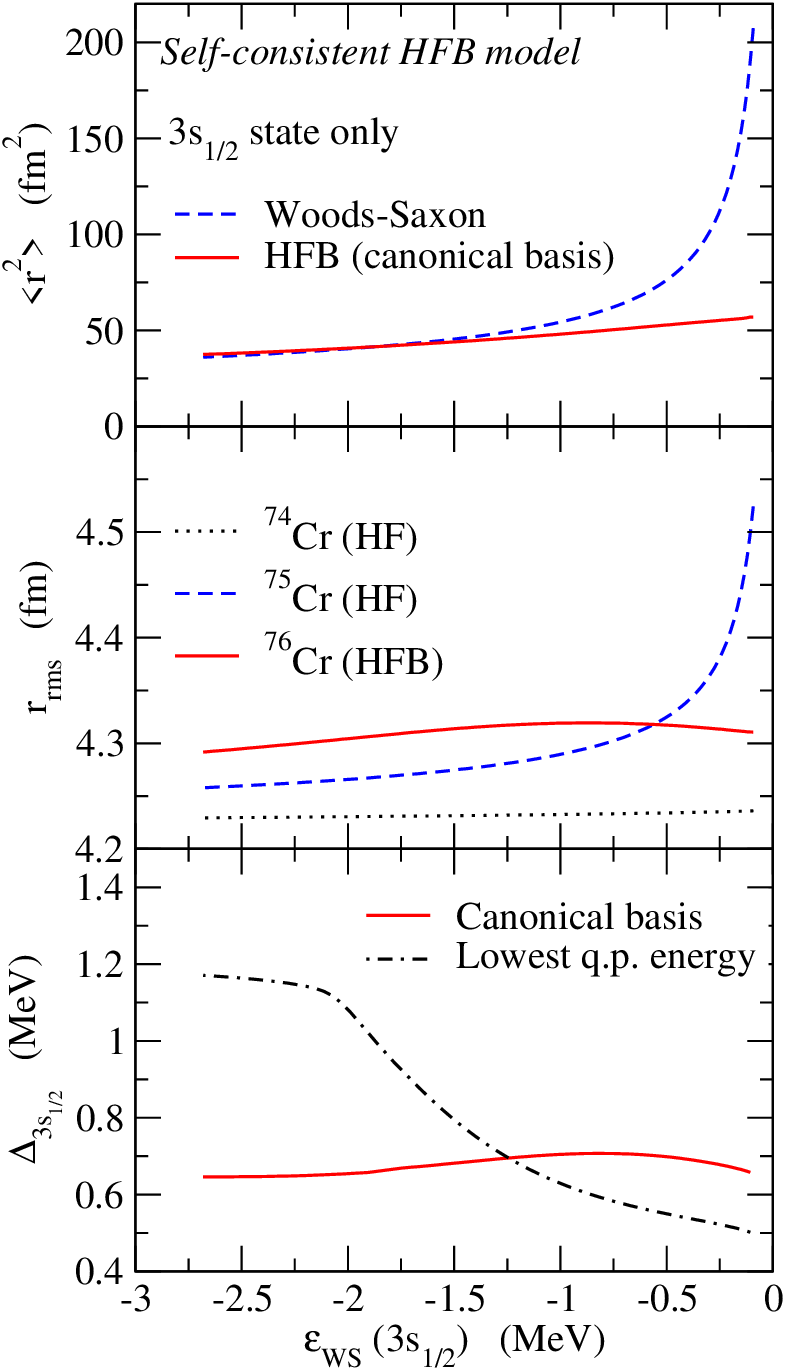}
\caption{(Color online) 
The mean square radii and the paring gap 
for $^{76}$Cr nucleus 
as a function of 
the single particle energy $\varepsilon_{\rm WS}$ 
for the 3s$_{1/2}$ orbit in a Woods-Saxon potential. 
The top panel shows the mean square radius of the 3s$_{1/2}$ wave function 
obtained with (the solid line) and without (the dashed line) 
the pairing correlation. 
The middle panel shows the root-means-square (rms) radii 
for $^{74}$Cr (the dotted line), $^{75}$Cr (the dashed line), 
and $^{76}$Cr (the solid line). 
These are obtained with the Hartree-Fock ($^{74}$Cr and 
$^{75}$Cr ) and the Hartree-Fock-Bogoliubov ($^{76}$Cr ) methods. 
The bottom panel shows the 
effective pairing gap for the 3s$_{1/2}$ state. The solid line 
is the pairing gap defined with the canonical basis, that is, the expectation 
value of the pair potential with respect to the canonical basis wave function 
for the 3s$_{1/2}$ state, while the dot-dashed line shows the lowest quasi-particle 
energy for the s$_{1/2}$ states.}
\end{figure}

The top panel of Fig. 1 shows the mean square radius of the 3$s_{1/2}$ state. 
In order to study the dependence on the binding energy, 
we vary the depth of the WS potential $V_0$ for neutron $s$ wave states while 
we keep the original value for the other angular momentum states. 
We also arbitrary change the single-particle energy for the 2d$_{5/2}$ state from $-0.38$ MeV 
to $-0.05$ MeV so that the 3s$_{1/2}$ state lies below the 2d$_{5/2}$ state. 
The dashed line is obtained with the single-particle wave function
for the mean-field Hamiltonian $\hat{h}$, 
while the solid line is obtained with the wave function for the canonical 
basis in the HFB calculations.  
The radius of the single-particle wave function 
for the $s$-wave state increases rapidly as the single-particle energy $\epsilon$ 
approaches zero, and eventually 
diverges 
in the limit of $\epsilon_{\rm WS}\rightarrow 0$. 
In contrast, 
the HFB wave function shows only a moderate increase of the radius 
even in the limit of $\epsilon_{\rm WS}\rightarrow 0$.    
The middle panel of Fig. 1 shows the root-mean-square (rms) radius for the whole 
nucleus, by taking into account 
the contribution of the other orbits as well. 
For comparison, we also show the rms radii for $^{74}$Cr and $^{75}$Cr 
obtained with the same mean-field potential but without including the pairing 
correlation. 
Notice that 
the rms radius of $^{76}$Cr is larger than that of $^{75}$Cr 
in the range of $\epsilon_{\rm WS}<-0.56$ MeV.  
This is due to the coupling of single-particle wave functions to 
a larger model space, including continuum, induced by the pairing correlations. 
On the other hand, 
in the limit of $\epsilon_{\rm WS}\rightarrow 0$, 
the rms radius of $^{75}$Cr shows a divergent feature 
while that of $^{76}$Cr is almost constant. 
for the 3s$_{1/2}$ state. The solid line 
shows the pairing gap evaluated with the canonical basis, $\Delta_i^{(\rm can)}$ for 
$i=3$s$_{1/2}$, while the dot-dashed line is the lowest quasi-particle energy $E_i$. 
Notice that the right hand side of Eq. (\ref{EHFB}) is simply a diagonal component of 
the HFB Hamiltonian in the canonical basis, while the left hand side is obtained by 
diagonalizing the HFB matrix\cite{DNW96}. Due to the off-diagonal components, 
Eq. (\ref{EHFB}) holds only approximately, and thus 
$E_i$ may be smaller than $\Delta_i^{\rm{(can)}}$ in actual calculations. 
One can see in the figure that the effective pairing gaps persist  
even in the limit of $\epsilon_{\rm WS}\rightarrow 0$, leading to 
the reduction of the radius of HFB wave function as 
is shown in the top panel of Fig. 1.

\begin{figure} 
\includegraphics[scale=0.5,clip]{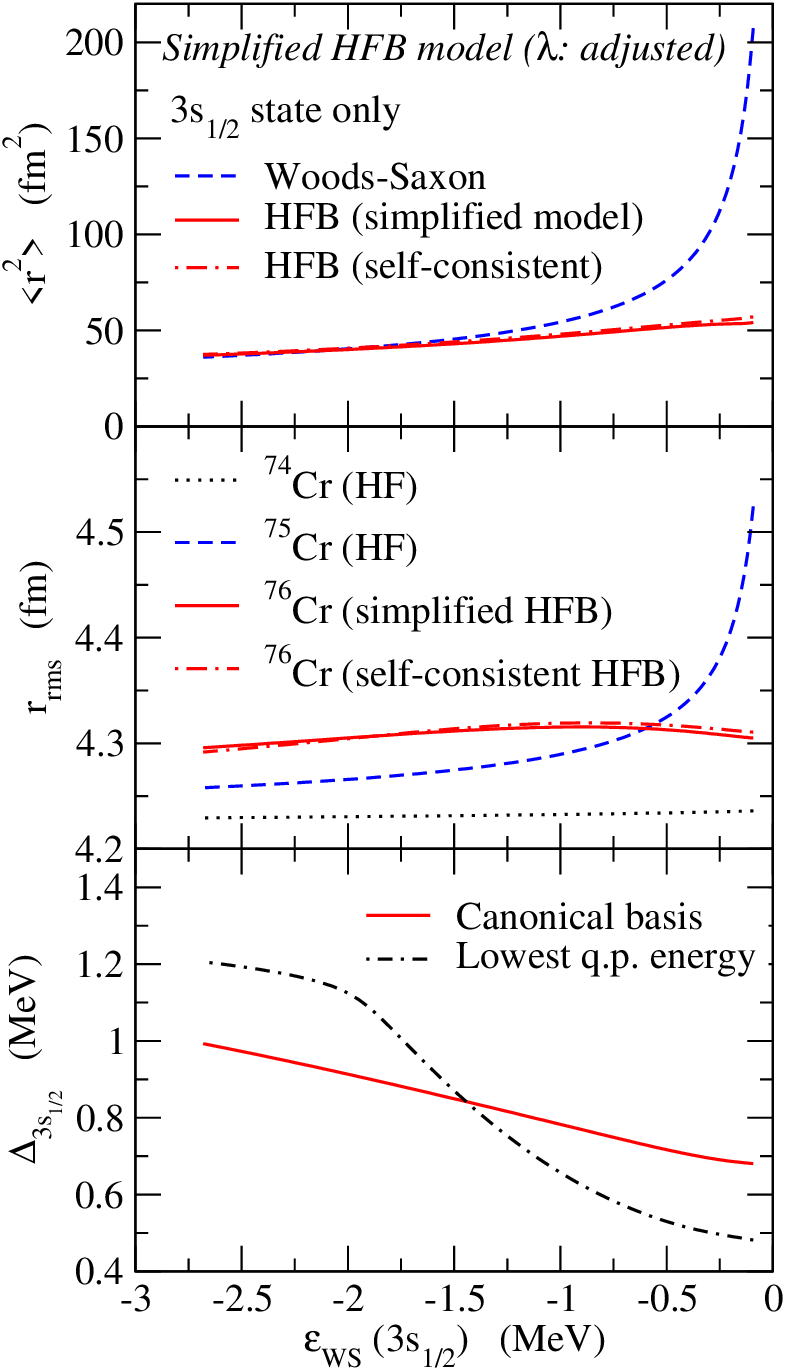}
\caption{(Color online) 
Same as Fig. 1, but with the simplified HFB model, in which the pair potential 
is assumed to be proportional to the derivative of the Fermi function. 
Only the Fermi energy $\lambda$ is determined self-consistently in this model. 
The dot-dashed lines in the top and the middle panels denote the results of 
the self-consistent calculations shown in Fig. 1. }
\end{figure}

In the simplified HFB model of Ref. \cite{Hamamoto}, it was claimed that the 
effective paring gap is diminished or quenched substantially for 
low $l$ orbits with  $l=0$ and 
 $l=1$.  In this model, the radial dependence of the 
pairing field $\Delta(r)$ is fixed either as a Fermi-type function (volume-type) 
or a derivative of the Fermi function (surface-type) with the same 
surface diffuseness parameter as in the mean-field potential. 
Furthermore, the Fermi energy is set equal to the single-particle energy, $\epsilon$, 
for the mean-field Hamiltonian $\hat{h}$, that is, $\lambda=\epsilon$. 
The effective pairing gap is defined in Ref. \cite{Hamamoto}
to be identical to the corresponding quasi-particle energy. If the energy in the canonical basis 
were the same as the single-particle energy, Eq. (\ref{EHFB}) indeed yields 
\be
E_i=\Delta_i^{(\rm can)}. 
\label{Delta-eff}
\ee
Care must be needed, however, since 
in general $\epsilon_i^{(\rm can)}$ deviates from the single-particle energy, 
$\epsilon_i$ in the HFB.  
Moreover, when the effective gap is plotted as a function of $\epsilon$ as has been 
done in Ref. \cite{Hamamoto}, 
setting $\lambda=\epsilon$ leads to a violation of particle number in this model, 
whose effect may be large in the limit of $\epsilon\to0$. 

\begin{figure} 
\includegraphics[scale=0.5,clip]{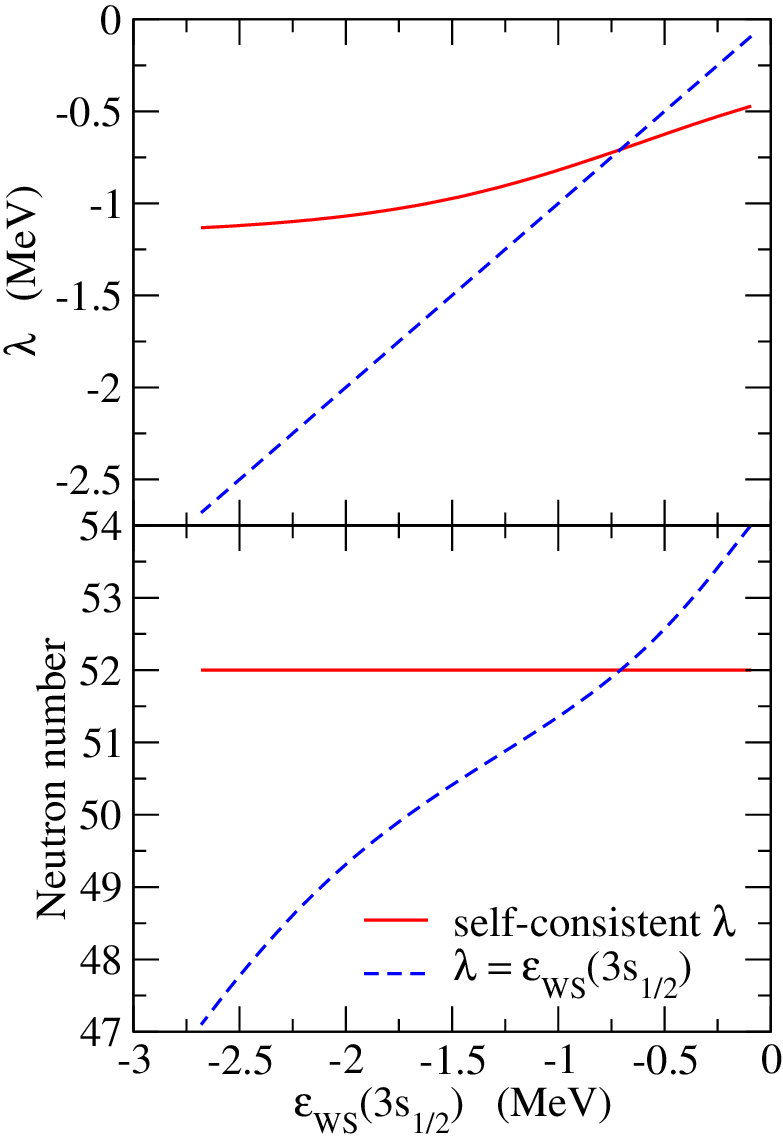}
\caption{(Color online) The Fermi energy $\lambda$ and the neutron number 
calculated with the simplified HFB model. 
These are plotted as a function of the single particle energy 
$\epsilon_{\rm WS}$ for the 3s$_{1/2}$ 
orbit in a Woods-Saxon potential. 
The solid lines are obtained by 
adjusting the Fermi energy so that the neutron number is $N$=52, while 
the dashed lines are obtained by setting $\lambda=\epsilon_{\rm WS}$. 
}
\end{figure}

In order to investigate the consistency of the simplified model of 
Ref. \cite{Hamamoto}, we repeat the same 
calculations shown in Fig. 1 by assuming that the 
pair potential $\Delta(r)$ is proportional to $r\cdot df/dr$, where $f(r)$ is given 
by Eq. (\ref{fermifunction}). 
We use the proportional constant of $-1.107$ MeV, that leads to the same value for the 
average pairing gap, 
\begin{equation}
\bar\Delta=\frac{\int^\infty_0r^2dr\,\Delta(r)\rho(r)}{\int^\infty_0r^2dr\,\rho(r)}, 
\end{equation}
as that in the self-consistent calculations shown in Fig. 1 for 
$\epsilon_{\rm WS}(3s_{1/2})=-0.257$ MeV. 
We keep this value in varying the depth of the Woods-Saxon potential, $-V_0$. 

We first keep the particle number to be a constant ($N$=52) 
and determine the Fermi energy self-consistently 
within this simplified model. Figure 2 show the results of such calculations. 
For comparison, the top and the middle panels also show by the dot-dashed lines 
the results of the self-consistent calculations, which have already been 
shown in Fig. 1. It is remarkable that this model yields a similar rms radius 
to that of  the self-consistent calculation. The effective pairing gaps show 
somewhat different behavior from those in the self-consistent calculation, especially 
for 
the pairing gap defined with the canonical basis (see the bottom panel). 
However, it 
should be emphasized that the pairing gaps stay finite in this calculation 
in the limit of vanishing 
single-particle energy, as in the self-consistent calculation shown in 
Fig. 1. 
This implies that the self-consistency for the pair potential 
is not important, as far as the rms radius is concerned.  

\begin{figure} 
\includegraphics[scale=0.5,clip]{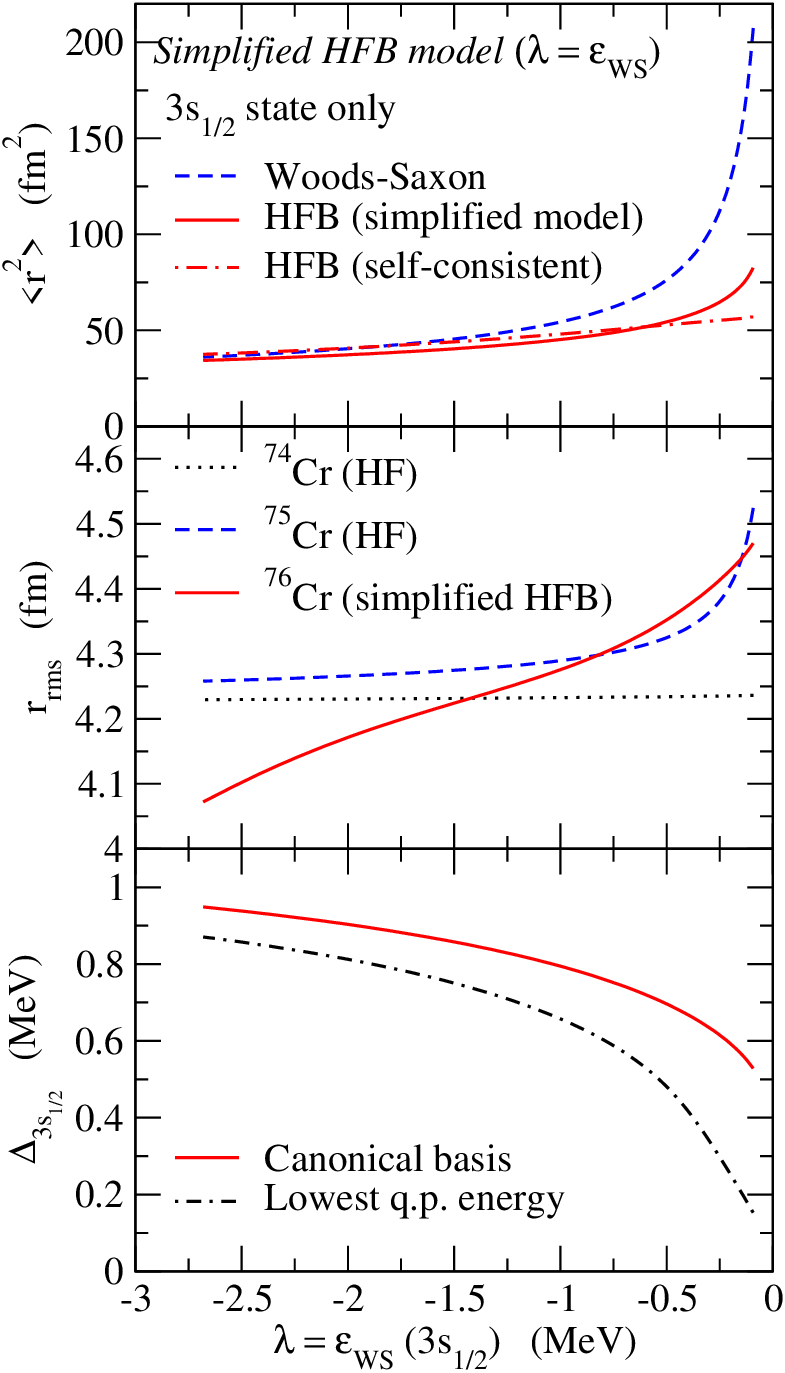}
\caption{(Color online) 
Same as Fig. 2, but by fixing the Fermi energy to be the same as 
the single-particle energy for the 3s$_{1/2}$ state, 
$\lambda=\epsilon_{\rm WS} (3s_{1/2})$. }
\end{figure}

We next carry out the calculation by setting the Fermi energy to be the same as 
the single-particle energy for the 3s$_{1/2}$ state, 
$\lambda=\epsilon (3s_{1/2})$. 
In this calculation, the number of particle changes as we vary the depth 
of the Woods-Saxon potential. 
The Fermi energy $\lambda$ and the neutron number are shown in 
the upper and the lower panels of Fig. 3, respectively, by the dashed lines. 
For comparison, the figure also shows the results of the previous 
calculation shown in Fig. 2, that is, those obtained by adjusting 
the Fermi energy so that the neutron number is a 
constant (see the solid lines). The variation of the particle number is 
large, that is, from 47 to 54 in the range of single-particle energy 
shown in Fig. 3. 
The radii and the effective pairing gaps are shown in Fig. 4. 
As one can clearly see, 
this non-self-consistent calculation yields considerably different 
results from the self-consistent calculations. Firstly, the 
reduction of the mean square radius of the single-particle orbit 
is somewhat underestimated, although the effect is still 
large (see the top panel). Secondly, the rms radius obtained with this model 
is completely inconsistent with the result of the self-consistent 
calculation as shown in 
the middle panel. Lastly, the effective pairing gaps drops off in the limit 
of $\epsilon\to 0$ (see the bottom panel). Particularly, the lowest quasi-particle 
energy is substantially dismissed, that is a similar behavior as that shown 
in Ref. \cite{Hamamoto}. Evidently, 
the claim of Ref. \cite{Hamamoto} that 
the pairing gap disappears 
in the zero energy limit is an artifact of 
setting $\lambda=\epsilon$. If the Fermi energy is determined self-consistently for a
given particle number, the effective pairing gap
persists even if the pairing potential 
is pre-fixed as shown in Fig. 2. 

\section{Odd-even staggering of reaction cross sections}

Let us now investigate how the pairing correlation affects 
the reaction cross sections of weakly-bound nuclei. 
To this end, we use the Glauber theory 
\cite{Glauber,BD04,OKYS01,OYS92,HSAK07,HSCB10}. 
In the optical limit approximation of the Glauber theory, 
the reaction cross section $\sigma_R$ can be calculated as 
\cite{OKYS01,OYS92,HSAK07,HSCB10},
\begin{equation}
\sigma_R=\int d\vec{b}\,\left(1-\left|e^{i\chi(\vec{b})}\right|^2\right),
\end{equation}
with 
\begin{equation}
i\chi(\vec{b})=-\int d\vec{r}d\vec{r}'\rho_P(\vec{r})\rho_T(\vec{r}')
\Gamma_{NN}(\vec{s}-\vec{s}'+\vec{b}).
\label{pshift}
\end{equation}
Here, $\rho_P$ and $\rho_T$ are the projectile and the target densities, 
respectively, and $\vec{b}$ is the impact parameter. $\vec{s}$ and $\vec{s}'$ 
are the transverse components of $\vec{r}$ and $\vec{r}'$, respectively, 
that is, $\vec{s}=(\vec{r}\cdot\vec{e}_b)\,\vec{e}_b$ and 
$\vec{s}'=(\vec{r}'\cdot\vec{e}_b)\,\vec{e}_b$, 
where $\vec{e}_b=\vec{b}/b$ is the unit vector parallel to $\vec{b}$. 
$\Gamma_{NN}$ is the profile function for the $NN$ scattering, for which we 
assume to take a form of \cite{OKYS01,OYS92,HSAK07,HSCB10},
\begin{equation}
\Gamma_{NN}(\vec{b})=\frac{1-i\alpha}{4\pi\beta}\,\sigma_{NN}\exp\left(
-\frac{b^2}{2\beta}\right), 
\label{Gamma_NN}
\end{equation}
with $\sigma_{NN}$ being the total $NN$ cross section. 

It has been known that the optical limit approximation 
overestimates reaction cross sections for weakly-bound nuclei 
\cite{BES90,TUKS92,AKT96,AKTT96,AIS00}. 
In order to cure this problem, Abu-Ibrahim and Suzuki have proposed 
modifying the phase shift function $\chi(\vec{b})$ in Eq. (\ref{pshift}) 
to \cite{AIS00} 
\begin{eqnarray}
&&i\chi(\vec{b})=
-\frac{1}{2}\int d\vec{r}\rho_P(\vec{r}) \nonumber \\
&&\times
\left[1-\exp\left(-\int d\vec{r}'\rho_T(\vec{r}')
\Gamma_{NN}(\vec{s}-\vec{s}'+\vec{b})\right)\right] \nonumber \\
&&-\frac{1}{2}\int d\vec{r}'\rho_T(\vec{r}') \nonumber \\
&&\times\left[1-\exp\left(-\int d\vec{r}\rho_P(\vec{r})
\Gamma_{NN}(\vec{s}'-\vec{s}+\vec{b})\right)\right].
\label{chi}
\end{eqnarray}
With this prescription, the effects of multiple scattering between a  
projectile nucleon and the target nucleus, and that between a  
target nucleon and the projectile nucleus, are included to some extent 
\cite{AIS00}. 

\begin{figure} 
\includegraphics[scale=0.5,clip]{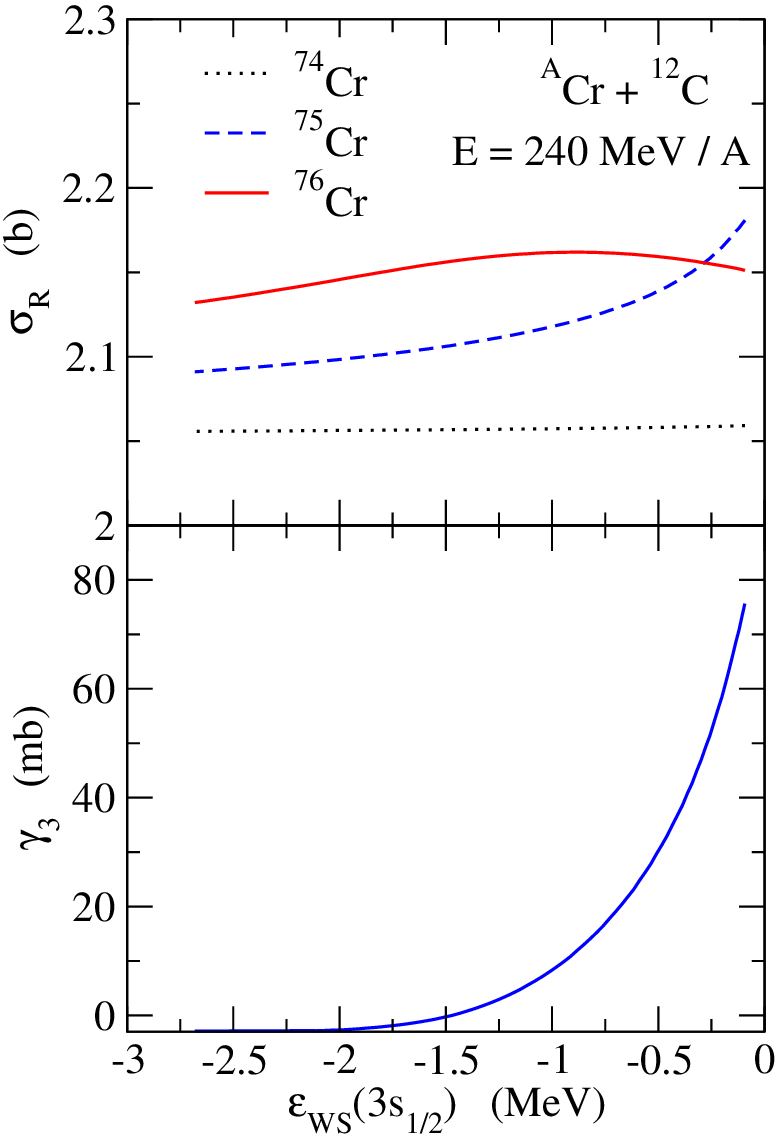}
\caption{(Color online) 
(The upper panel) 
Reaction cross sections for $^{74,75,76}$Cr + $^{12}$C reactions 
at $E$=240 MeV/nucleon, obtained with the modified optical limit approximation 
of the Glauber theory, Eq. (\ref{chi}). The density distributions for the 
$^{74,75,76}$Cr are constructed from the HFB calculations shown in Fig. 1. 
(The lower panel) 
The staggering parameter defined by Eq. (\ref{EQ-gam}). 
}
\end{figure}

The upper panel of Fig. 5 
shows the reaction cross sections for $^{74,75,76}$Cr + $^{12}$C 
reactions at $E$=240 MeV/nucleon, obtained with the phase shift function 
given by Eq. (\ref{chi}). For the density of the Cr isotopes, we use 
the results of the HFB calculations shown in Fig. 1. 
For the density of the target nucleus $^{12}$C, we use the same density 
distribution as that given in Ref. \cite{OYS92}. 
In the actual calculation, we treat the proton-neutron and the 
proton-proton/neutron-neutron scattering separately and 
use the parameters given in Table I in Ref. \cite{AIHKS08} 
for the profile function $\Gamma_{NN}$. 
In order to evaluate the phase shift function, we use the two-dimensional 
Fourier transform technique \cite{BS95}. 
We give its explicit form in the Appendix. 
The reaction cross sections shown in Fig. 5 show a similar behavior as in the 
rms radii shown in the middle panel of Fig. 1, as is expected. That is, 
the reaction cross sections for $^{76}$Cr and $^{75}$Cr are inverted at 
a small biding energy, due to the pairing effect shown 
in Fig. 1.  This leads to a large odd-even staggering in reaction 
cross sections for weakly bound nuclei\cite{HS11}. 

\begin{figure} 
\includegraphics[scale=0.4,clip]{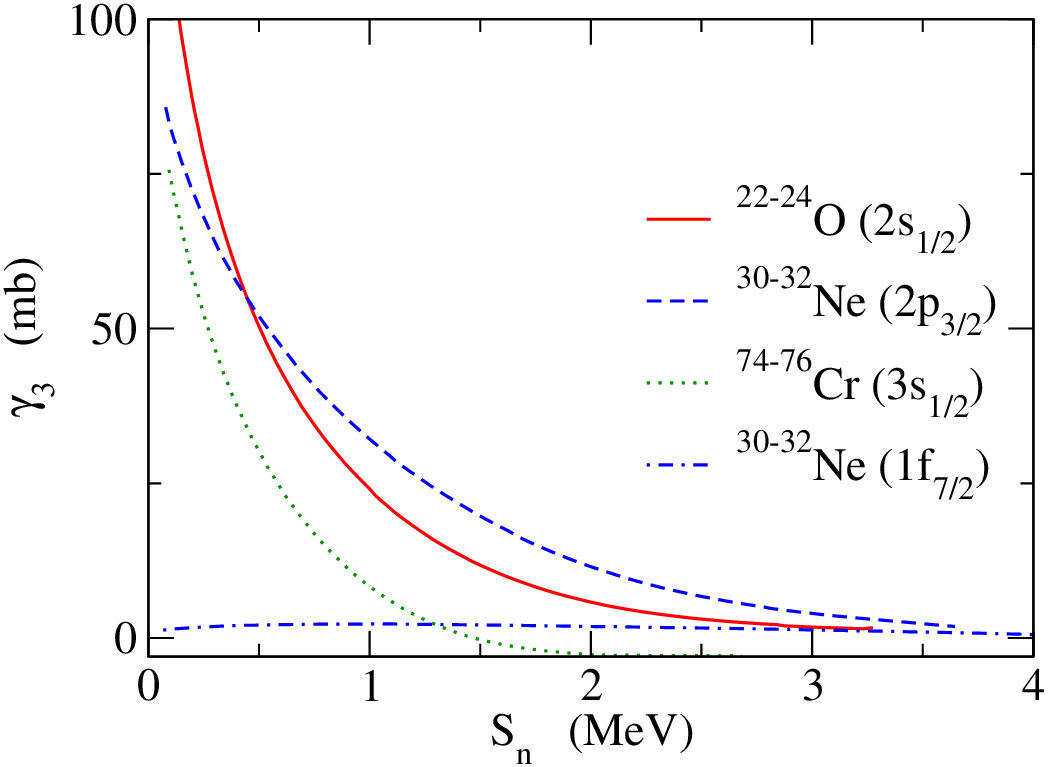}
\caption{(Color online) 
The staggering parameter $\gamma_3$ defined by Eq. 
\eqref{EQ-gam} as a function of 
the neutron separation energy S$_n$ for the odd-mass nuclei.  
The solid and the dotted lines correspond 
to the reaction cross sections for $^{22,23,24}$O+$^{12}$C 
and $^{74,75,76}$Cr+$^{12}$C at $E$=240 MeV/nucleon, respectively. 
The dashed and the dot-dashed lines show the results for 
$^{30,31,32}$Ne+$^{12}$C at $E$=240 MeV/nucleon, in which the valence neutron 
in $^{31}$Ne occupies the 2p$_{3/2}$ or the 1f$_{7/2}$ orbits, respectively. 
}
\end{figure}

In order to quantify the OES of reaction cross sections, we 
introduce the staggering parameter defined by 
\be
\gamma_3=(-)^{A}\frac{\sigma_R(A+1)-2\sigma_R(A)+\sigma_R(A-1)}{2},
\label{EQ-gam}
\ee
where $\sigma_R(A)$ is the reaction cross section of a nucleus
 with mass number $A$.  We can define the 
same quantity also for rms radii. 
Notice that this staggering parameter 
is similar to the one often used for the OES of binding 
energy, that is, the pairing gap \cite{SDN98,DMNSS01,BBNSS09}. 
The lower panel of Fig. 5 shows the staggering parameter 
$\gamma_3$ for the $^{74,75,76}$Cr nuclei as 
a function of the single-particle energy, $\epsilon_{\rm WS}$. 
One can clearly see that the staggering parameter $\gamma_3$ 
increases rapidly for small separation energies, and goes up to 
a large value reaching $\gamma_3\sim$ 80 mb.  

In order to find out
a general trend of the staggering parameter, Fig. 6 shows 
the value of $\gamma_3$ for various orbits with 
2s$_{1/2}$, 3s$_{1/2}$, 2p$_{3/2}$ and 1f$_{7/2}$. The values for the 
2s$_{1/2}$ and 2p$_{3/2}$ orbits correspond to the reaction cross sections 
for the $^{22,23,24}$O and $^{30,31,32}$Ne nuclei, respectively, 
calculated in Ref. \cite{HS11}. 
The value for the 1f$_{7/2}$ orbits corresponds to the reaction 
cross sections for $^{30,31,32}$Ne nuclei, obtained with the diffuseness 
parameter of the mean-field Woods-Saxon potential of 
$a$=0.65 fm. 
One can clearly see that $\gamma_3$ 
for the low $l$ orbits with $l=0$ and 
$l=1$ show a rapid increase at small separation energies, 
the $l=0$ orbits increasing more rapidly than the $l=1$ orbit. 
In contrast, 
the high $l$ orbit with $l=3$ does not show any anomaly 
in the limit of $\epsilon\to 0$. 
These features are quite similar to the growth of a halo structure only 
in the low  $l$ orbits due to a zero or small 
centrifugal barrier. 

\begin{figure} 
\includegraphics[scale=0.4,clip]{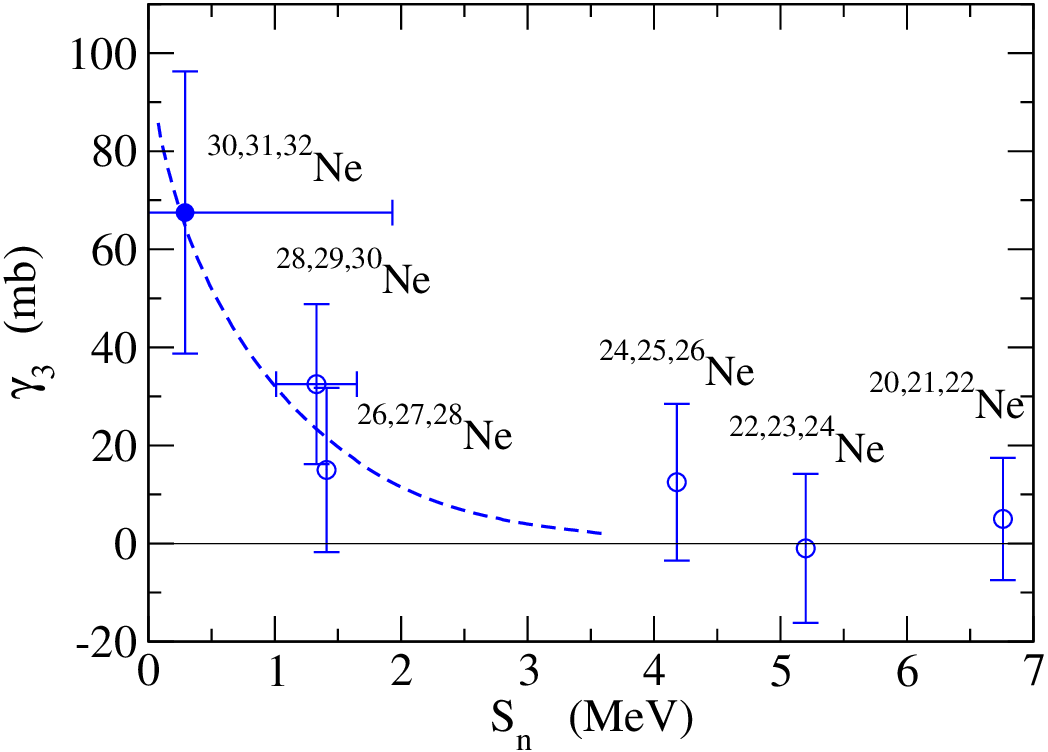}
\caption{(Color online) 
The experimental staggering 
parameter $\gamma_3$ of reaction cross sections 
defined by Eq. \eqref{EQ-gam} 
for the Ne isotopes with the $^{12}$C target at $E$=240 MeV/nucleon.  
This is plotted 
as a function of the neutron separation energy S$_n$ of the odd-$A$ nuclei.  
The experimental data 
for the reaction cross sections are taken from Ref. \cite{Takechi10}, while 
the empirical separation energies are taken from Refs. \cite{Audi,J07}. 
The dashed line is the calculated staggering parameter for 
the $^{30,31,32}$Ne isotopes, assuming that the valence neutron in 
of 
$^{31}$Ne occupies the 2p$_3/2$ orbit. 
}
\end{figure}

The experimental staggering parameters $\gamma_3$ are plotted in Fig. 7 
for Ne isotopes as a function of the neutron separation energy for the 
odd-mass nuclei. We use the experimental reaction cross sections given 
in Ref. \cite{Takechi11} while we evaluate 
the separation energies with 
the empirical binding energies listed in Ref. \cite{Audi}. 
For the neutron separation energy for the $^{31}$Ne nucleus, we use the value 
in Ref. \cite{J07}. The experimental uncertainties of the staggering parameter 
are obtained as
\begin{equation}
\delta\gamma_3=\frac{\sqrt{(\delta\sigma_R(A+1))^2+4(\delta\sigma_R(A))^2
+(\delta\sigma_R(A-1))^2}}{2},
\end{equation}
where $\delta\sigma_R(A)$ is the experimental uncertainty for the 
reaction cross section of a nucleus with mass number $A$. 
The figure also shows by the dashed line 
the calculated staggering parameter for the 
$^{30,31,32}$Ne nuclei with the 2p$_{3/2}$ orbit, that has 
been shown also in Fig. 6. 
One sees that the 
experimental staggering parameter 
agrees with the calculated value for 
$^{30,31,32}$Ne nuclei when one assumes that the valence neutron in 
$^{31}$Ne occupies the 2p$_{3/2}$ orbit. 
Furthermore, although the structure of lighter odd-A Ne isotopes is not 
known well, it is interesting to see that 
the empirical staggering parameters closely follow the 
calculated values for the 2p$_{3/2}$ orbit. 
This may indicate that the low-$l$ single-particle 
orbits are appreciably mixed in 
these Ne isotopes due to the deformation effects\cite{H10,UHS11}.

\section{Summary}

We have studied the odd-even staggering (OES) of 
the reaction cross sections 
by using the Hartree-Fock-Bogoliubov model. To this end, 
we have introduced the staggering parameter $\gamma_3$ 
defined with a 3 point difference formula in order to 
clarify the relation between 
the magnitude of OES and the neutron separation energy. 
We have shown that 
the OES parameter increases largely for low-$l$ orbits with $l=0$ and 
$l=1$ at small separation energies. 
The experimental staggering parameter for the Ne isotopes 
show a similar increase. 
On the other hand, we have found that 
the staggering parameter stays almost a constant value, 
$\gamma_3\sim 2$ mb for higher $l$ orbits with {\it e.g.,} $l=3$.  
The increase of $\gamma_3$ is 
induced by the finite pairing correlations in 
the zero separation energy limit.  
In this respect, we have shown that the effective pairing gap for 
the 3s$_{1/2}$ orbit in the $^{76}$Cr nucleus persists even in the limit of 
vanishing separation energy. 
This remains the same even if the pair potential is prefixed, as long as 
the chemical potential is adjusted to keep the particle number to be the same. 
We have shown that such simplified HFB model well reproduces 
the results of the 
self-consistent HFB model for the root-mean-square radius. 

The staggering parameter proposed in this paper provides a good measure 
for the OES of reaction cross sections. Further systematic experimental 
studies would be helpful in order to clarify the pairing correlations in 
weakly-bound nuclei and in the limit of zero neutron separation energy. 

\begin{acknowledgments}
We would like to thank M.  Takechi for fruitful discussions on 
the experimental data. 
We thank also P. Schuck and C.A. Bertulani for useful discussions. 
This work was supported by the Japanese
Ministry of Education, Culture, Sports, Science and Technology
by Grant-in-Aid for Scientific Research under
the program numbers  (C) 22540262 and  20540277.
\end{acknowledgments}

\appendix

\section{Evaluation of phase shift function with 
the Fourier transform method}

In this paper, we evaluate the phase shift functions given by Eqs. 
(\ref{pshift}) and (\ref{chi}) using the Fourier transform technique 
\cite{BS95}. 
First we notice that Eq. (\ref{pshift}) can be expressed as  
\begin{equation}
i\chi(\vec{b})=-\int d\vec{s}d\vec{s}'\rho^{(z)}_P(\vec{s})\rho^{(z)}_T(\vec{s}')
\Gamma_{NN}(\vec{s}-\vec{s}'+\vec{b}), 
\end{equation}
with 
\begin{equation}
\rho^{(z)}_P(\vec{s})\equiv \int dz \,\rho_P(\vec{s},z), 
~~~
\rho^{(z)}_T(\vec{s}')\equiv \int dz \,\rho_T(\vec{s}',z). 
\end{equation}
The two-dimensional Fourier transform of $i\chi(\vec{b})$ in Eq. 
(\ref{pshift}) then reads 
\begin{eqnarray}
i\tilde{\chi}(\vec{q})&=& \int d\vec{b}\,i\chi(\vec{b})
\,e^{i\vec{q}\cdot\vec{b}}, \\
&=&
-\int d\vec{b}\,e^{i\vec{q}\cdot(\vec{b}+\vec{s}-\vec{s}')}
\,\Gamma_{NN}(\vec{b}+\vec{s}-\vec{s}') \nonumber \\
&\times&
\int d\vec{s}\,e^{-i\vec{q}\cdot\vec{s}}\,\rho^{(z)}_P(\vec{s})
\int d\vec{s}'\,e^{i\vec{q}\cdot\vec{s}'}\,\rho^{(z)}_T(\vec{s}'), \\
&=&-\tilde{\Gamma}_{NN}(\vec{q})(\tilde{\rho}_P^{(z)}(\vec{q}))^*
\tilde{\rho}_T^{(z)}(\vec{q}),
\label{2DFT}
\end{eqnarray}
where $\tilde{\Gamma}_{NN}, \tilde{\rho}_P^{(z)}$, and $\tilde{\rho}_T^{(z)}$ 
are the two-dimensional Fourier transform of 
${\Gamma}_{NN}, {\rho}_P^{(z)}$, and ${\rho}_T^{(z)}$, respectively. 
For the profile function given by Eq. (\ref{Gamma_NN}), its Fourier transform 
reads,
\begin{equation}
\tilde{\Gamma}_{NN}(q)=\frac{1-i\alpha}{4\pi\beta}\,\sigma_{NN}
\cdot 2\beta^2\pi \exp\left(
-\frac{\beta^2q^2}{2}\right). 
\end{equation}
The Fourier transform of the density distribution is evaluated as 
\begin{equation}
\tilde{\rho}^{(z)}(\vec{q})=
\int dzd\vec{s}\,e^{i\vec{q}\cdot\vec{s}}\,\rho(\vec{s},z)=
\tilde{\rho}(\vec{Q}),
\end{equation}
where $\tilde{\rho}(\vec{Q})$ is the three-dimensional Fourier transform 
of the density at $\vec{Q}=(\vec{q},Q_z=0)$. 
For a spherical density, $\rho(r)$, 
$\tilde{\rho}(\vec{Q})$ depends only on $|\vec{Q}|=q$, that is, 
\begin{equation}
\tilde{\rho}^{(z)}(\vec{q})
=\tilde{\rho}(q)=4\pi\int^\infty_0r^2dr\,\rho(r)j_0(qr),
\end{equation}
where $j_0(qr)$ is the spherical Bessel function of zero-th order. 
Taking the inverse Fourier transform of Eq. (\ref{2DFT}), 
the phase shift function is calculated as 
\begin{eqnarray}
i\chi(\vec{b})
&=&-\int\frac{d\vec{q}}{(2\pi)^2}\,
\tilde{\Gamma}_{NN}(q)\tilde{\rho}_P(q)\tilde{\rho}_T(q)e^{-i\vec{q}\cdot\vec{b}}, \\
&=&-\int^\infty_0\frac{qdq}{2\pi}\,
\tilde{\Gamma}_{NN}(q)\tilde{\rho}_P(q)\tilde{\rho}_T(q)J_0(qb), 
\end{eqnarray}
where $J_0(qb)$ is the Bessel function of zero-th order. 
A similar technique has been used to evaluate a double folding 
potential in heavy-ion reactions \cite{SL79,BS97,KOBO07}. 

One can apply the same method to evaluate the phase shift function 
given by Eq. (\ref{chi}). First notice that 
\begin{eqnarray}
\Gamma_{NT}(\vec{b})&\equiv& 
1-\exp\left(-\int d\vec{r}'\rho_T(\vec{r}')
\Gamma_{NN}(\vec{b}-\vec{s}')\right), \\
&=&1 -
\exp\left(-\int^\infty_0 \frac{qdq}{2\pi}\,\tilde{\Gamma}_{NN}(q)\tilde{\rho}_T(q)J_0(qb)\right), \nonumber \\
\end{eqnarray}
depends only on $b=|\vec{b}|$. 
This leads to 
\begin{eqnarray}
i\chi(b)&=&
-\frac{1}{4\pi}\int^\infty_0qdq\,\tilde{\Gamma}_{NT}(q)\tilde{\rho}_P(q)J_0(qb) \nonumber \\
&&-\frac{1}{4\pi}\int^\infty_0qdq\,\tilde{\Gamma}_{NP}(q)\tilde{\rho}_T(q)
J_0(qb), 
\end{eqnarray}
where $\tilde{\Gamma}_{NT}(q)$ is calculated as 
\begin{equation}
\tilde{\Gamma}_{NT}(q)=2\pi\int^\infty_0bdb\,\Gamma_{NT}(b)J_0(qb).
\end{equation}

\end{document}